\documentstyle[psfig]{l-aa}

\def\aeta{A\&A }

\def\apj{ApJ }

\def\mn{MNRAS }

\def\gsim{\lower.4ex\hbox{$\;\buildrel >\over{\scriptstyle\sim}\;$}}
\def\lsim{\lower.4ex\hbox{$\;\buildrel <\over{\scriptstyle\sim}\;$}}

\begin{document}
%
%
\thesaurus{11.09.4;11.17.1;11.17.4 Q0528-250;11.08.1}
\title{
Molecules in the $z_{\rm abs}$~=~2.8112 damped system toward 
PKS~0528--250\thanks{Based on observations
collected at the European Southern Observatory, La Silla, Chile}
}
\author{R. Srianand$^1$ \and Patrick Petitjean\inst{2,3}}
\institute{$^1$IUCAA, Post Bag 4, Ganesh Khind, Pune 411 007, India  \\
$^2$Institut d'Astrophysique de Paris -- CNRS, 98bis Boulevard 
Arago, F-75014 Paris, France\\
$^3$UA CNRS 173 -- DAEC, Observatoire de Paris-Meudon, F-92195 Meudon
Cedex, France \\
}
\date{ }
\offprints{R. Srianand}
\maketitle
\markboth{}{}
\begin{abstract}
We present a detailed analysis of a high resolution spectrum of the
damped Ly$\alpha$ system at $z_{\rm abs}$~=~2.8112 
toward PKS~0528-250. The absorption redshift is slightly larger than
the emission redshift of the quasar.
We estimate the column density of H$_2$ molecules 
$N$(H$_2$)~$\sim$~6$\times$10$^{16}$~cm$^{-2}$ and the fractional 
abundance of H$_2$, $f$~=~5.4$\times$10$^{-5}$. 
The excitation
temperature derived for different transitions suggests that the
kinetic temperature of the cloud is $\sim$200~K
and the density $n$~$\sim$~1000~cm$^{-3}$. The cloud therefore
has a dimension of $\sim$1~pc along the line of sight. Since
it obscures the broad-line emission region, its transverse dimension should
be larger than 10~pc.\par
We obtain upper
limits on the column densities of C~{\sc i} ($<$~10$^{12.7}$~cm$^{-2}$)
and CO ($<$~10$^{13.2}$~cm$^{-2}$; $N$(CO)/$N$(H~{\sc
i})~$<$~7$\times$10$^{-9}$). 
We suggest that the
ratio $N$(H$_2$)/$N$(C~{\sc i}) is a useful indicator of the 
physical conditions in the absorber. 
Simple photo-ionization models assuming solar relative abundances 
show that radiation fields with spectra similar to typical AGNs or
starbursts are unable to reproduce all the constraints and in particular
the surprisingly small $N$(C~{\sc i})/$N$(H$_2$) and
$N$(Mg~{\sc i})/$N$(H$_2$) ratios. 
In view of the models we explored, the most likely ionizing spectrum
is a composite of a UV-"big bump" possibly produced by a local starburst
and a power-law spectrum from the QSO that provides the X-rays.
Dust is needed to explain the production of molecules in the cloud. The
amount of dust is broadly consistent with the [Cr/Zn] abundance
determination. 


\end{abstract}


\section{Introduction} \label{intr}

QSO absorption line systems probe the baryonic matter over most of the
history of the Universe (0~$<$~$z$~$\la$~5).  The so-called damped
Ly$\alpha$ (hereafter DLA) systems are characterized by a very large
H~{\sc i} column density ($N$(H~{\sc i})~$\ga$~2$\times$10$^{20}$
~cm$^{-2}$), similar to the one usually seen through local spiral disks. The
case for these systems to be produced by proto-galactic disks is
supported by the fact that the cosmological density of gas associated
with these systems is of the same order of magnitude as the
cosmological density of stars at present epochs (Wolfe 1996).  Moreover
the presence of heavy  elements ($Z \sim 0.1 ~ Z_\odot$) and
the redshift evolution of metallicity suggest the ongoing star
formation activities in these systems (Pettini et al.  1997), while
strong metal line systems have been demonstrated to be associated with
galaxies at low and intermediate $z$ (e.g.  Bergeron \& Boiss\'e 1991).
It has also been shown that the profiles of the lines arising in the
neutral gas show evidence for rotation (Wolfe 1996, Prochaska \& Wolfe
1997).  Whether these arguments are enough to demonstrate that DLA
systems arise in large disks is a matter of debate however.  Indeed
simulations have shown that the progenitors of present day disks of
galaxies could look like an aggregate of well separated dense clumps at
high redshift. The kinematics could be explained by relative motions of
the clumps with very little rotation (Haehnelt et al. 1997, Ledoux et al.
1998). Moreover,
using {\sl HST} high spatial resolution images of the field of seven
quasars whose spectra contain DLA lines at intermediate redshifts
(0.4~$\la$~$z$~$\la$~1), Le~Brun et al. (1996) show that, in all cases,
at least one galaxy candidate is present within 4~arcsec from the
quasar. There is no dominant
morphological type in their sample: three candidates are spiral
galaxies, three are compact objects and two are amorphous low surface
brightness galaxies. Therefore, although the nature of the DLA systems is
unclear they trace the densest regions of the Universe where
star formation occurs.\par
It is thus surprising that despite intensive searches, the amount of
H$_2$ molecules seems quite low in DLA systems in
contrast to what is observed in our own galaxy. Two detections of H$_2$
molecules in high redshift DLA systems have been reported.  Recently Ge
\& Bechtold (1997) have found strong absorptions in the $z_{\rm
abs}$~=~1.9731 DLA system toward Q~0013--004. They derive
$N$(H$_2$)~=~6.9$\times$ 10$^{19}$~cm$^{-2}$, $b$~=~15~km~s$^{-1}$,
$T_{\rm ex}$~$\sim$~70~K and $n$(H)~$\sim$~300~cm$^{-3}$ for a total
hydrogen column density $N$(H)~=~6.4$\times$10$^{20}$~cm$^{-2}$. This
system has by far the largest H$_2$ abundance
$f$~=~2$N$(H$_2$)/[2$N$(H$_2$)~+~$N$(H~{\sc i})] $\sim$~0.22$\pm$0.05
observed in high z DLA systems. However the exact number should be confirmed 
using a higher resolution data. Other searches have led to much smaller
values or upper limits ($f$~$<$~10$^{-6}$, Black et al. 1987, Chaffee
et al. 1988, Levshakov et al. 1992). Levshakov \& Varshalovich (1985)
suggested that H$_2$ molecules could be present in the $z_{\rm abs}$~=~2.8112
system toward PKS~0528--250. This claim has been confirmed by Foltz et
al. (1988) using a 1~\AA~ resolution spectrum. The latter authors
derive $N$(H$_2$)~=~10$^{18}$~cm$^{-2}$, $b$~=~5~km~s$^{-1}$, $T_{\rm
ex}$~=~100~K and log~$N$(H~{\sc i})~=~21.1$\pm$0.3. By fitting the
damped absorption together with the Ly$\alpha$ emission from the
quasar, M\o ller \& Warren (1993) find log~$N$(H~{\sc i})~=~21.35.
Three Ly$\alpha$ emission-line objects have been detected within
100$h^{-1}$~kpc from the quasar by M\o ller \& Warren (1993) and
confirmed by Warren \& M\o ller (1996) to have redshifts within
200~km~s$^{-1}$ from the redshift of the DLA system
($z_{\rm abs}$~=~2.8112 as measured on the Ni~{\sc ii} lines by Meyer \&
York 1987). The widths of the Ly$\alpha$ emission lines are very large
($>$~600~km~s$^{-1}$) and continuum emission could be present (Warren
\& M\o ller 1996); this suggests that the gas is not predominantly
ionized by the quasar and that star-formation may occur in the clouds,
a conclusion reached as well by Ge et al. (1997). The proximity of the
quasar makes the case difficult however and careful analysis is
needed.\par
In this paper we use much higher spectral resolution data to reanalyze the
molecular lines in this system. We present the observations in
Section~2, the results  in Section~3 and discuss the low $N$(C~{\sc
i})/$N$(H$_2$) ratio inferred in Section~4.

\section{Observations} \label{s2}
The observations were carried out at the F/8 Cassegrain focus of the
3.6~m telescope at La Silla, ESO Chile. The spectra were obtained with
the ESO echelle spectrograph (CASPEC) and long camera.  Three
exposures of 5400~s each were obtained under good seeing conditions in
November 1995.  A 300 line~mm$^{-1}$ cross disperser was used in
combination with a 31.6 line~mm$^{-1}$ echelle grating.  The detector
was a Tektronix CCD with 1124$\times$1024 pixels of 15~$\mu$m square
and a read--out noise of 2.5 electrons. For each exposure on the object,
flat field images and wavelength comparison Thorium--Argon spectra were
recorded. The slit width was 1.6~arcsec corresponding to a spectral
resolution of $R$~$\sim$~36000 over the wavelength range
3650~--~4850~\AA. The spectra were binned in the direction
perpendicular to the dispersion axis.  The accuracy in the
wavelength calibration measured on the calibrated Thorium--Argon
spectra is about 0.03~\AA. \par The data were reduced using the echelle
reduction package provided by IRAF.  The cosmic--ray events have been
removed in the regions between object spectra before extraction of the
object.  The exposures were co--added to increase the signal to noise
ratio. During this merging procedure the cosmic--ray events affecting
the object pixels were recognized and eliminated.  The background sky
spectrum was difficult to extract separately due to the small spacing
between the orders in the blue.  Instead, we have carefully fitted the
zero level to the bottom of the numerous saturated lines in the
Ly$\alpha$ forest. The uncertainty on the determination can be
estimated to be 5\%.  \par 
We identify all the absorption features with
equivalent widths larger than 5$\times$FWHM$\times$$\sigma$ where
$\sigma$ is the noise rms in the adjacent continuum. We use the
line list to identify the metal line systems in the spectra and
more specifically all the absorptions associated with
the $z_{\rm abs}$~=~2.8112 DLA system.\par
\section{Results } \label{s3}
\subsection {Low ionization metal lines}
\begin{table}
\label{Tabab}
\caption[]{Heavy element lines from the damped system}
\begin{tabular}{cccc}
\hline
\multicolumn {1} {c} {ION}&  
\multicolumn {1} {c} {\rm log N} &
\multicolumn {1} {c} {[Z/H]}& Ref\\
\hline
H{\sc~i}    & 21.35$\pm$0.10 &....          & 1\\
C{\sc~i}    & $<12.77       $&....          & 3\\
Ar{\sc~i}   & 14.46$\pm$0.23 &$-1.43\pm0.63$& 3\\
Mg{\sc~i}  & $<$12.8         &              & 2,3\\
Mg{\sc~ii}  & $<15.88       $&$<-1.04      $& 2\\
Si{\sc~ii}  & 16.00$\pm$0.04 &$-0.90\pm0.11$& 2\\
S{\sc~ii}   & 15.27$\pm$0.06 &$-1.03\pm0.04$& 2\\
Cr{\sc~ii}  & 13.65$\pm$0.12 &$-1.38\pm0.16$& 2\\
Fe{\sc~ii}  & 15.46$\pm$0.09 &$-1.41\pm0.05$& 3\\
Ni{\sc~ii}  & 13.89$\pm$0.03 &$-1.71\pm0.10$& 2\\
Zn{\sc~ii}  & 13.09$\pm$0.07 &$-0.91\pm0.12$& 2\\
\hline
\multicolumn{4}{l}{$^1$ M\o ller \& Warren 1993; $^2$ Lu et al. 1996;
$^3$ This work} \\
\end{tabular}
\end{table}
\begin{figure}
\psfig{figure=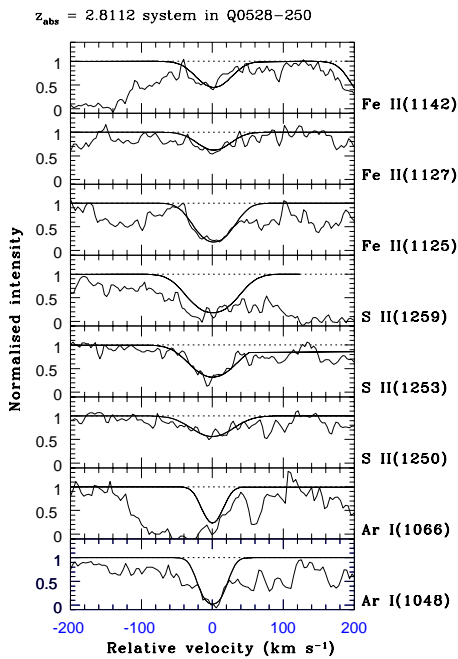,height=9.cm,angle=0}
\caption[]{Fit result for few weak heavy element transitions in
the $z_{\rm abs}$ = 2.8112 damped system}
\end{figure}
Lu et al.
(1996) have obtained column densities of most of the elements in the
DLA system considered here.
In addition to the lines identified by Lu et al (1996) we find a
possible Ar~{\sc i} doublet from this molecular cloud (see Fig.~1). Since 
our spectrum extends further into the blue, we could also
obtain a better estimate of the Fe~{\sc ii} column density using the
lines from very weak transitions (e.g. 
Fe~{\sc ii}$\lambda$1125, Fe~{\sc ii}$\lambda$1127, 
Fe~{\sc ii}$\lambda$1142 etc., see Fig.~1). We also obtained an upper limit
to the C~{\sc i} column density which will be useful for constraining the
model parameters. The column densities of different species 
are given in Table~1. Since in the neutral phase of DLA systems the 
dominant ionization state of the elements listed in Table~1 is
either the neutral (for argon) or the singly ionized (for the others) state, 
we can estimate element abundances. 
The values we obtain for the S~{\sc ii}
column density is about a factor of two less than that of Lu et al.
The reason is that we detect and fit only the strongest component whereas 
Lu et al. integrate the apparent optical depth from $v$~=~--80 to
300~km~s$^{-1}$. For consistency of the discussion, we adopt the 
value from Lu et al. Subsequently, since singly ionized species can be 
found well outside the region where hydrogen is neutral, it must be 
noticed that the values of the ratios $N$(X$^{+}$)/$N$(H~{\sc i}) are 
upper limits.

An interesting limit on the Mg~{\sc i} column density
can be obtained from the Lu et al. data. Mg~{\sc i}$\lambda$2026 is
blended with Zn~{\sc ii}$\lambda$2026. However from Fig.~9 of
Lu et al., we can see that the optical depth of the line is smaller
than 0.1. If we consider, as for Zn~{\sc ii} and Cr~{\sc ii} that
the absorption is spread over 40~km~s$^{-1}$ we can infer that
$w_{\rm r}$~$<$~0.027~\AA. Using the oscillator strength given by
Morton(1991), we find 
$N$(Mg~{\sc i})~$<$~6.6$\times$10$^{12}$~cm$^{-2}$.

Formation of H$_2$ molecules is most efficient in the presence of dust
grains (see Section~3.2). In DLA systems the amount of dust 
is usually estimated from the ratio $N$(Zn~{\sc ii})/$N$(Cr~{\sc ii}) assuming
that, like in the ISM of our galaxy, Zn is not heavily depleted onto
dust grains whereas Cr is depleted
(Meyer et al. 1989; Pettini et al. 1994, 1997; Lu et al. 1996).  
Some controversy has arisen however about the presence of dust
in DLA systems.
Lu et al. (1996) have argued that nucleosynthesis alone can explain 
the  element abundance ratios observed in 
DLA systems and that the presence of dust is thus questionable
whereas Pettini et al. (1997) have claimed that even though 
nucleosynthesis history of the gas must play a role, the consistent 
depletion levels of the refractory elements indicates the presence of
dust. We discuss the abundance ratios of Table~1 considering the two
possibilities in turn.

It is known that Zn and S are not heavily depleted into dust grains
(Sembach \& Savage 1996) and should reflect the gas-phase abundances
which is thus of the order of 0.1--0.2~$Z_{\odot}$. 
The average value of [Cr/Zn] in the diffuse interstellar clouds with 
molecular fraction log~$f$(H$_2$) between --6 and --4 is --1.18$\pm$0.53
(Roth \& Blades 1995). The observed [Cr/Zn]~=~--0.47 in
the $z$~=~2.8112 system suggests that about 70\% of Cr is depleted 
onto dust grains in the damped system compared to 99\% 
in the diffuse interstellar medium of our galaxy. 
The observed
Ni~{\sc~ii}/Cr~ {\sc ii} in the 2.8112 system is consistent with the
ratio in the interstellar medium. Silicon has the same gas-phase abundance
as zinc (see Table~1)
which is consistent with the much lower depletion of silicon as 
compared to iron, chromium and nickel in the interstellar medium.
The non detection of Mg~{\sc~i} is intriguing. Although the upper limit
on Mg~{\sc ii}
is consistent with the other element abundances, some depletion could
be needed (see Section~4). All these are consistent with the presence of 
dust in the system with small departure of the total element abundance 
ratios from the solar values.

It is interesting to note however that in our galaxy, stars with 
[Fe/H]~$\sim$~-1, have [Si/Fe]~$\sim$~[S/Fe]~$\sim$~0.4 (e.g. Fran\c cois 1987)
and [Cr/Fe]~$\sim$~[Ni/Fe]~$\sim$~0 (Magain 1989, Gratton \& Sneden 1991)
which is approximately what we observe. Thus the assumption of depletion
onto dust grains may be questionable.
The upper limit on magnesium is consistent with [Mg/Fe]~$\sim$~0.4
(Magain 1989). However a better limit on this element would definitively
be very helpful. However if we accept the gas-phase metallicities as
the true metallicities, then the observed value [Zn/Fe]~=~0.5 is not 
consistent with the
value observed in the same stars ([Zn/Fe]~=~0, Sneden \& Crocker 1988).

In the case of the $z_{\rm abs}$~=~2.8112 system toward PKS~0528--250,
we can investigate further this issue  
by asking the question whether dust is needed to produce the 
observed amount of H$_2$ molecules (see Section~4).

\subsection {{\rm H}$_2$ molecules}
The consequence of the spectral resolution being nine times larger than
the previous published data can be appreciated by comparing Fig.~2 with
Fig.~4 of Foltz et al. (1988). In our data, a large number of molecular
lines are left unblended making the determination of the line
parameters much reliable.\par

We have derived the $N$(H$_2$) and $b$ values using a Voigt profile
fitting code (Khare et al. 1997). The oscillator strengths of different
transitions given in Morton \& Dinerstein (1976) have been used.  Only
the clean and unblended lines from each rotational level are fitted.
The best fit is obtained with a single component model although a few
discrepancies between the fit and the data may argue for a
two-component model (see also the comment by Jenkins \& Peimbert 1997
about this system). The resulting fit is shown in Fig.~2 for a few
transitions. The fit parameters are given in Table~2 where the second,
third and fourth columns give the statistical weights of different
levels,  the column density and the number of lines used in the model
respectively.  The best value for the velocity dispersion is $b$ =
9.1~km~s$^{-1}$.
\begin{table}
\label{Tabh2}
\caption[]{$H_2$ column densities and excitation temperatures}
\begin{tabular}{cccccc}
\hline
\multicolumn {1} {c} {LEVEL}& \multicolumn {1} {c} {g}& 
\multicolumn {1} {c} {${\rm N_J }$} &
\multicolumn {1} {c} {Number}&
\multicolumn {1} {c} {${\rm T_{ex}}$}&
\multicolumn {1} {c} {$\rm J_{ex}$}\\
\multicolumn {1} {c} { } & \multicolumn {1} {c} { }&
\multicolumn {1} {c} {${\rm (10^{16}~cm^{-2})}$} &
\multicolumn {1} {c} { }&
\multicolumn {1} {c} { (K)}&
\multicolumn {1} {c} { }\\
\hline
J = 0& 1 & 0.965$\pm$0.330 & 4 & ..... & .....\\
J = 1& 9 & 4.110$\pm$0.009 & 7 & 229$\pm$95 & 0-1\\
J = 2& 5 & 0.417$\pm$0.063 & 8 & 208$\pm$36 & 0-2\\
J = 3& 21& 0.216$\pm$0.029 & 3 & 225$\pm$08 & 0-3\\
     &   &                 &   & 224$\pm$09 & 1-3\\
J = 4& 9 & $\le 0.03$      &...& $\le 300$  & 0-4\\
\hline
\end{tabular}
\end{table}

The total column density $N$(H$_2$) in the cloud is 6$\times
10^{16}~{\rm cm}^{-2}$ and the measured fractional H$_2$ abundance is
$f$~=~2~$N$(H$_2$)/[2~$N$(H$_2$)~+~$N$(H~{\sc i})] = 5.4$\times$10$^{-5}$. This
value is obtained using $N$(H~{\sc i}) = 2.2 $\times 10^{21}{\rm cm^{-2}}$
(M\o ller \& Warren 1993), and assuming that all the neutral hydrogen
contributing to the damped Ly~$\alpha$ absorption originates from the
same region as H$_2$. If this were not the case, $f$ would be higher.  
The molecular fraction obtained here is two orders of
magnitude less than the values quoted by Foltz et al. (1988). However
as noted by the latter authors, the column density is very sensitive to
the Doppler parameter and their value should be considered as an upper
limit. 

In the case of thermal equilibrium, the ratio of column densities
of two rotational levels are given by the Boltzman equation,

\begin{equation}
{{N(J)\over N(0)}\; =\; {g_J\over g_0}\; exp \bigg[\;-{{BJ(J+1)}\over 
kT_{ex}}\; \bigg]}.
\end{equation}

where $g_J$ is the statistical weight of any level $J$ in the vibrational
ground state, $T_{\rm ex}$ is the excitation temperature and the
constant $B/k$ is $\simeq$ 85 K. The values of $T_{\rm ex}$ derived for
different transitions are given in column 5 of Table~2.  
Lines from higher "$J$" levels (i.e. $J \ge$4) are weak and we could derive
only upper limits to column densities.

It is believed that the $J$ = 1 level is mostly populated by collisions
so the excitation temperature, ${T_{01}}$, is roughly equal to the
kinetic temperature. As can be seen from Table~2, the excitation
temperatures derived for different levels are similar, suggesting that
$T_{\rm ex}$ indeed equals the kinetic temperature which is therefore
close to $\sim$~220~K in this molecular cloud.

\begin{figure}
\centerline{\vbox{
\psfig{figure=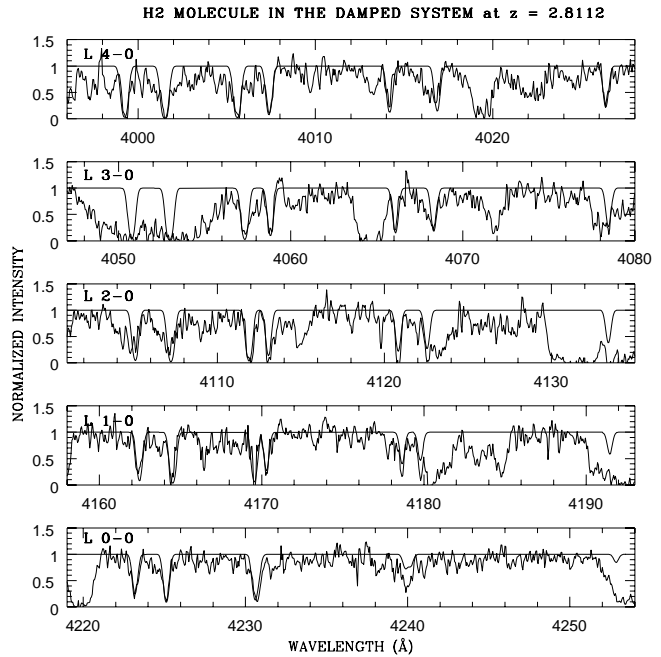,height=9.cm,angle=0}
}}
\caption[]{Fit results for a few rotational transitions of 
the H$_2$ Lyman absorption
bands in the $z_{\rm abs}$ = 2.8112 damped system .
}
\end{figure}

As suggested by Jura (1975) the column density ratio
$N(2)/N(0)$ and $N(3)/N(1)$ are controlled by the kinetic temperature
when $n$(H)$>$20 and $n$(H)$>$300~cm$^{-3}$ respectively. 
Our limit on the population of the J~=~4 rotational level 
do not add much constraint on the density. However, Ge \& Bechtold (1997) 
report a new determination $N$(4)/$N$(0)~=~4.5$\times$10$^{-3}$
by Songaila \& Cowie (1995). This new value implies an excitation temperature
$T_{04}$~=~224~K. The calculations by Nishimura (1968) indicate
that all this is consistent with a kinetic temperature
of $T$~$\sim$~220~K and a density $n$~$\sim$~1000~cm$^{-3}$.




Using the asymptotic approximation given by Jura(1974) for
the ratio of the molecular to neutral hydrogen densities, 
$n$(H$_2$)/$n$(H), as a function of the optical depth within a molecular
cloud, we can derive an estimate of the photo-dissociation rate of molecules,
I, due to the incident UV flux,

\begin{equation}
I={nR\over \delta}\sqrt{N\over 2<f>}
\end{equation}
where $\delta$~=~4.2$\times$10$^{5}$, $n$ is the hydrogen density, 
$R$ is the rate of molecule formation onto dust grains; $N$ is the total
H~{\sc i} column density and $<$f$>$ is the mean molecular fraction
observed through the cloud. If we scale $R$ with the metallicities from
the value observed in the ISM, then we can estimate that
$R$~$\sim$~3$\times$10$^{-18}$~cm$^{3}$s$^{-1}$. This may well be an
upper limit since the dust to gas ratio in the damped system is
smaller than in the ISM (see Secion 3.1). Using the density and the
molecular faction  derived above (i.e., $n$~$\sim$~1000~cm$^{-3}$ 
and $f$~$\sim$~5.4$\times$10$^{-5}$ ) we estimate the photodissociation
rate of molecules , $I$~$\sim$~3.2$\times$10$^{-8}$~s$^{-1}$, and the 
photo-absorption rate in the Lyman and Werner bands, 
$\beta_{\rm o}$~$\sim$~$I$/0.11~$\sim$~3$\times$10$^{-7}$~s$^{-1}$ (Jura 1975).
This large value results from the large H~{\sc i} column density 
albeit small molecular fraction seen in $z_{\rm abs}$~=~2.8112 damped system.
We are confident that $f$, $n$ and the metallicities are known within a 
factor of two.

The estimated value of the photo-absorption rate,
$\beta_{\rm o}$, is 3$\times$10$^{-7}$~s$^{-1}$. This value is an
order of magnitude higher than the values obtained by Ge \& Bechtold
(1997).
It is also two orders of magnitude higher than the average value
for the interstellar radiation field in our galaxy (i.e. $\beta_{\rm o} =
5\times$10$^{-10}$~s$^{-1}$). The average value for the intergalactic
medium radiation field calculated for $J_{21}$(912 \AA) = 1, is 
$\beta_{\rm o}
= 2\times$10$^{-12}$~s$^{-1}$. Thus the radiation field in the
$z$~=~2.8112 absorbing cloud is not dominated by the
intergalactic UV radiation field. The most probable sources are either
the radiation from the quasar or from the young stars in the molecular
cloud itself.

If we take quasar alone (see however Section~4) to be the ionizing
source
then, using $m_{\rm B}=18.17$, $q_{\rm o}=0.5$, $H_{\rm o}=75~{\rm
km~s^{-1}~Mpc^{-1}}$, the optical to UV spectral index $\alpha=-1$
($f_\nu\propto \nu^{-\alpha}$) and the upper limit on $\beta_{\rm o}$, we
estimate that the H$_2$ cloud should be at a distance larger than
10~kpc from the quasar.  


If we assume that the H$_2$ lines are thermally broadened then the velocity
dispersion of 9 km~s$^{-1}$ gives a kinetic temperature of $\sim$~2000~K. 
This is higher than the kinetic temperature inferred from the
excitation temperatures. Thus it is possible that there are more than
one component in the H$_2$ absorption. This may introduce a small
error in the total H$_2$ column density. However the lines are not
heavily saturated and hence the errors cannot be large.

\subsection{CO molecules}
In the galactic diffuse clouds,
$N$(CO)~$\sim$~10$^{-7}$$\times$$N$(H). As the high
redshift DLA systems have, on average, one tenth of
the solar metallicity we would expect the $N$(CO)/$N$(H) ratio to be
lower than 10$^{-7}$. Indeed optical and radio searches (e.g.
Levshakov et al. 1989; Wilkind \& Combes 1994
and references therein) fail to detect CO both in emission and
absorption in high $z$ damped absorbers.

However CO emission is detected from a few high redshift quasars
and ultra-luminous infrared galaxies (see e.g. Barvainis et al. 1994,
Ohta et al. 1996; Omont et al. 1996; Guilloteau et al. 1997). 
These detections suggest that a 
large amount of mass in host galaxies (or surrounding regions) of some of the
high redshift AGNs is in the form of molecules. 

We have searched for the presence of absorption due to  CO A-X
molecular bands in a $FWHM$ = 0.6~\AA~ spectrum obtained by Petitjean
\& Bergeron (1994) primarily for studying intervening C~{\sc iv} systems (see
their paper for the observational details).  No strong CO absorption is
detected from the $z$~=~2.8112 system (see Fig.~3).\par
\begin{figure}
\centerline{\vbox{
\psfig{figure=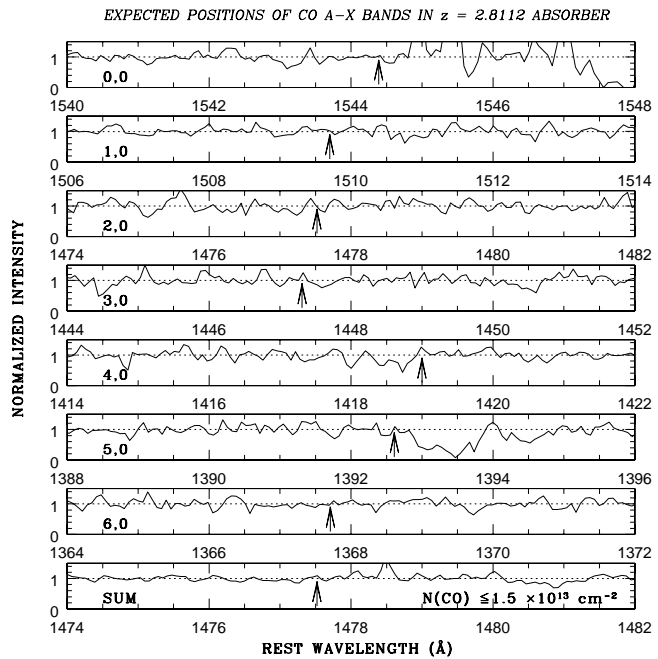,height=9.cm,angle=0}}}
\caption[]{Plot showing the spectrum in the rest wavelength of 
different A-X bands in CO due
the $z_{\rm abs}$ = 2.8112 damped system. The centres of each band
is marked with an arrow.}
\end{figure}
In order to improve the signal to noise we followed the stacking method
prescribed by Levshakov et al. (1989). The resulting combined spectrum
at the rest frame of the 2-0 band is given in the last panel of the
Fig.~3.  No clear line is seen at the expected position.  Using the
weighted oscillator strength of all the bands and a velocity dispersion
of 9 km s$^{-1}$, similar to H$_2$, we obtain a 3$\sigma$ upper limit
for the column density of $N$(CO)  $\simeq 1.5 \times 10^{13}\; {\rm
cm^{-2}}$. This limit is consistent with other determinations (Ge et
al. 1997).  This is a factor of two higher than the upper limits
obtained for the damped systems along the line of sight to PHL 957 and
1331+170 (Levshakov et al. 1989). This limit leads to $N$(CO)/$N$(H)$\le
7\times 10^{-9}$.

Federman et al. (1980) have estimated the CO column density along the
lines of sight towards 48 bright stars. The results of their analysis
suggest that CO is readily detected when $N$(H$_2$)~$>$~10$^{19}{\rm
cm^{-2}}$.  Thus the non detection of CO in the $z$ = 2.8112 system is
consistent with the $N$(CO)/$N$(H$_2$) ratios seen in the diffuse
interstellar medium.  They also noted that $N$(C~{\sc i}) is, on
average, 10 times larger than $N$(CO). If $N$(C~{\sc i}) is greater than
$N$(CO) in the system under consideration also, then the upper limit on
$N$(C~{\sc i}) suggests that the value of $N$(CO) should be an
order of magnitude less than the derived upper limit.

\section{The low $N$(C~{\sc i})/$N$(H$_2$) ratio}

The ratios $N$(C~{\sc i})/$N$(H~{\sc i})~ $<$~2.2~10$^{-9}$ 
and $N$(H$_2$)/$N$(H~{\sc i}) ~=~2.7~10$^{-5}$ 
derived from this 
study are surprisingly low. Indeed in the interstellar medium of our Galaxy,
all the clouds with log~$N$(H~{\sc i})~$>$~21 have 
log~$N$(H$_2$)~$>$~19 and log~$N$(C~{\sc i})~$>$~14
(Jenkins \& Shaya 1979). 
It thus may be interesting to investigate the
reasons for such low ratios.

Formation of H$_2$ is expected on the surface of dust grains if the gas
is cool, dense and mostly neutral, and from the formation of negative
hydrogen if the gas is warm and dust free (see e.g. Jenkins \& Peimbert
1997). Given the neutral hydrogen column density of this system and the
evidence for depletion onto dust grains (see however Warren \& M\o ller
1996), the first process should dominate.  
Destruction is mainly due to UV photons.  The effective
photodissociation of H$_2$ takes place in the energy range 11.1--13.6
eV, through Lyman-Werner band line absorption.
%
%
Since the ionization potential of  C~{\sc i} is 11.2 eV, it is sensitive
to the same photo-ionization field as H$_2$ in the neutral phase. 
%
%
If the H$_2$ lines are saturated then the cloud is self-shielded
against further photodissocation of H$_2$. In this case however
large values of $N$(H$_2$) to $N$(C~{\sc i}) ratio are easily produced. 

%
\begin{figure}
\centerline{\vbox{
\psfig{figure=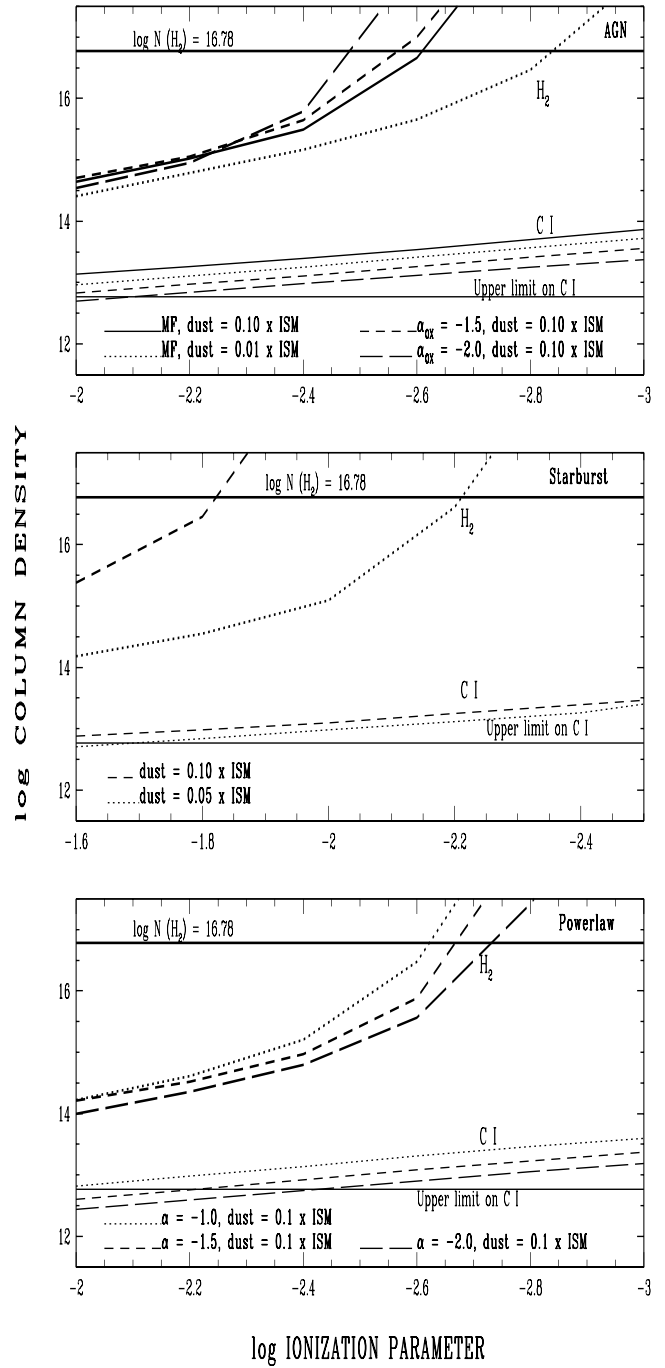,height=20.cm,width=9cm,angle=0}
}}
\caption[]{H$_2$ and C~{\sc i} column densities versus ionization
parameter for different ionizing spectra (the SEDs are given
in Fig.~\ref{sed}).
The cloud is modeled as a slab of constant density $n$~=~100~cm$^{-3}$
illuminated on one side by an incident flux, resulting in an ionization
parameter $U$ (ratio of the density of ionizing photons to the density
of hydrogen). Metallicity is a tenth of solar and the amount of dust is
taken as a fraction of the ISM dust content.
}
\label{models}
\end{figure}
\begin{figure}
\centerline{\vbox{
\psfig{figure=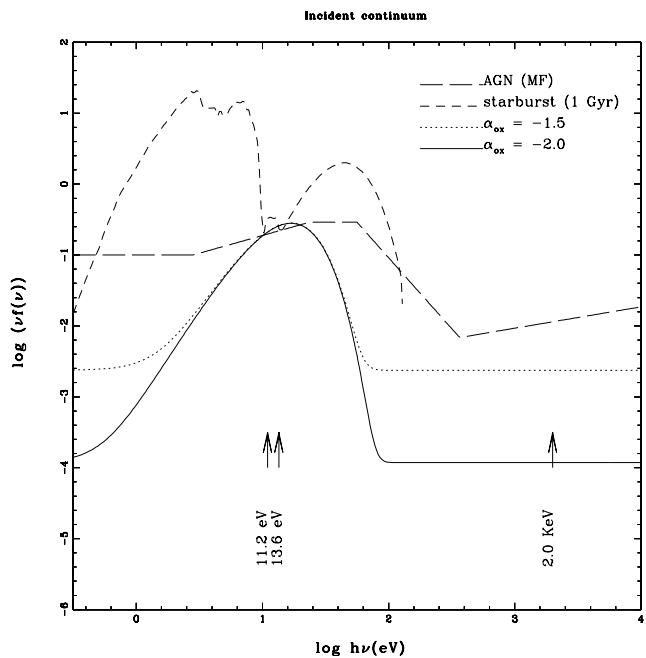,height=9.cm,angle=0}
}}
\caption[]{Spectral energy distributions of the incident flux 
irradiating the cloud from the models presented in the text.
The small-dashed curve corresponds to a 1~Gyr old starburst;
the long-dashed curve is the AGN spectrum described by Mathews 
\& Ferland (1987); the solid and dotted lines correspond to
power-law spectra with UV big-bumps, modeled as a black-body 
of temperature 50~000~K.}
\label{sed}
\end{figure}
We run grids of photoionization models using CLOUDY (Ferland 1993).
These models take into account photo-dissociation of H$_2$
molecules using the approximations outlined by Tielens and 
Hollenbach (1985).
The incident radiation field is averaged  
over the wavelength interval 912--1109~\AA~  and is redistributed 
according
to the interstellar radiation spectrum given by Habing (1968).

In our calculations we take the gas phase abundances to be 1/10 of 
solar.
The grain composition is assumed to be similar to that of the Galactic
ISM and we use the dust to gas ratio as a free parameter.  We perform
the photoionization calculations for different ionizing radiation
spectra and a range of ionizing parameters, $U$.  We consider four
possible spectral energy distribution for the incident radiation field
which are (1) a simple power--law of index $\alpha$,
$F(\nu)$~$\propto$~$\nu^{-\alpha}$; (2) a power-law with a "big bump"
modeled as a black-body of temperature 50~000~K, the relative
contributions of the power-law spectrum and the black-body is
characterized by the index $\alpha_{\rm
ox}$~=~log~$F$(2500~\AA)/$F$(2~keV); (3) a typical spectrum of
AGNs (Mathews \& Ferland 1987); and (4) a starburst (Charlot, private
communication). The optical to X-ray spectral energy distribution of
PKS~0528--250 is not well defined. Wilkes et al. (1994) failed to detect
X-ray emission from this quasar in their Einstein IPC data.  They
derive an upper limit on $\alpha_{\rm ox}$ of $-$1.34. It is known
that a typical quasar will have $\alpha_{\rm ox}$ in the range $-$1.0 to
$-$2.0 with a mean of $-$1.5 (e.g. Wilkes et al. 1994). 
Thus in our model we use either $\alpha_{\rm ox}$~=~$-$1.5 or $-$2.0.

The predicted column densities of H$_2$ and C~{\sc i} as a function of
the ionizing parameter $U$ for different models 
are given in Fig.~\ref{models} (the SEDs are shown for direct
comparison in Fig.~\ref{sed}). For all models, 
log~$N$(H~{\sc i})~=~21.35 and 
$n_{\rm H}$~=~100~cm$^{-3}$. We have checked that 
the exact value of $n_{\rm H}$ has little effect on the results.

To reproduce the small $N$(C~{\sc i})/$N$(H~{\sc i}) ratio, the
spectrum must be steep to favor ionization of C~{\sc i}. This is why a
power-law spectrum with $\alpha$~=~--2, a "big-bump" spectrum with
$\alpha_{\rm ox}$~=~--2 and the starburst spectrum can reproduce the
ratio with somewhat different ionizing parameters. The "big-bump"
spectrum is favored because the X-ray tail maintains the temperature of
the neutral gas above 100~K. The other models imply too low a
temperature compared to what we derived from the H$_2$ features.
However all the models fail to reproduce at the same time the H$_2$
column density. It
is apparent from the figure that steeper spectra could meet the
constraints.  However again the temperature would be too low. Moreover,
all the models predict log~$N$(Mg~{\sc i})~$>$~13.5. This is a factor
of seven larger than the upper limit we derived from the Lu et al.
(1996) data.  In the framework of these models, the only solution we
can find is to decrease the abundances of carbon by a factor of two and
that of magnesium by a factor of seven. This could reflect true
relative abundances or depletion onto dust grains.

Dust is needed to explain the H$_2$ column density as is apparent from
Fig.~\ref{models}. The H$^-$ process is more
efficient than the dust process at temperature $T$~$>$~500~K and when
$n_{\rm e}$ is not negligible, thus in the partially ionized region.
However the very large H~{\sc i} column density of this damped
Ly$\alpha$ system requires most of the gas to be neutral where this
process is negligible because of the low temperature resulting from the
steep spectrum needed to explain the lack of C~{\sc i} and Mg~{\sc i}.
\section{Conclusion } \label{s4}
By
fitting the different H$_2$ transitions that we detect in a
high resolution spectrum of PKS~0528--250,
we derive log~$N$(H$_2$)~$\sim$~16.78 and 
$T_{\rm ex}$~$\sim$~200~K that is most certainly the true kinetic temperature
of the gas. For this temperature, the relative populations of rotational levels
0 to 4 indicate that the density is of the order of 1000~cm$^{-3}$.
Therefore the dimension of the neutral cloud along the line
of sight is less than 1~pc. It must be noticed that the damped
absorber must cover the broad line region. Indeed there is
no residual flux in the bottom of the Ly$\alpha$ absorption
line which completely absorbs the Ly$\alpha$ emission from the
quasar over more than 5000~km~s$^{-1}$ (M\o ller \& Warren 1993).
Since the dimension of the BLR in QSOs can be 
approximated as $R$~$\sim$~0.3~$L_{46}^{0.5}$ where $L_{46}$ is
the bolometric luminosity in units of 10$^{46}$~erg~s$^{-1}$
(Collin, private communication). The radius
of the BLR in PKS~0528--250 is thus of the order of 10~pc. The transverse
dimension of the damped cloud must be thus larger than 10~pc
and the cloud must be quite flat. 

We can derive an upper limit on the
transverse dimension of the cloud by interpreting the non-dection
of redshifted 21 cm absorption (Carilli et al. 1996) as an effect
of partial covering factor of the continuum radio emission by the cloud.
Indeed the size of the radio source is 1~arcsec or 5$h^{-1}_{75}$~kpc.
If the spin temperature is equal to the kinetic temperature we
derived in sect. 3.2,
this implies that the covering factor should be less than 0.3 and
thus the radius of the cloud  along the transverse direction is 
less than 1~kpc. 


We determine upper limits for the C~{\sc i}, Mg~{\sc i} and CO
column densities
(log~$N$(C~{\sc i})~$<$~12.7, log~$N$(Mg~{\sc i})~$<$~12.8
and log~$N$(CO)~$<$~13.2). Based on simple photoionization models we
conclude that (i) no simple model can reproduce
at the same time the low $N$(C~{\sc i})/$N$(H~{\sc i}), 
$N$(Mg~{\sc i})/$N$(H~{\sc i}) 
ratios and the presence of
molecules at the level observed; 
(ii) steep spectra are required to reproduce
the low $N$(C~{\sc i})/$N$(H~{\sc i}) and $N$(Mg~{\sc i})/$N$(H~{\sc i}) 
ratios but they predict temperatures
smaller than 100~K in conflict with the excitation temperature derived
from the H$_2$ transitions; (iii) the only way to keep the temperature as 
high as 200~K is to allow for some X-ray flux heating the gas; (iv)
in the framework of these models, dust is needed to produce the
observed amount of molecules; (v) the gas phase abundances of
carbon and especially magnesium should be smaller than 0.1 of solar.

In view of the models we explored, the most likely ionizing spectrum
is a composite of a UV-"big bump" possibly produced by a local starburst
and a power-law spectrum from the QSO that provides the X-rays.


\begin{acknowledgements} 
We thank Drs. L. Cowie and J. Bechtold for useful information, St\'ephane
Charlot for providing us with the new models of starburst SED and
an anonymous referee for insightful comments.
\end{acknowledgements}

\end{document}